\documentclass[aps,pra,preprint,superscriptaddress,amsmath,showpacs,tightenlines,nofootinbib]{revtex4-1}

\usepackage[ngerman,english]{babel}
\makeatletter\AtBeginDocument{\let\@elt\relax}\makeatother
\usepackage{amsmath}
\DeclareMathOperator{\Tr}{Tr}
\usepackage{slashed}
\usepackage[utf8]{inputenc}
\usepackage{fontenc}
\usepackage{braket}
\usepackage{bbm}
\usepackage{cancel}
\usepackage{graphicx}
\usepackage[outdir=./]{epstopdf}
\usepackage{calrsfs}
\usepackage{emptypage}
\usepackage{fancyhdr}
\usepackage[pdfstartview=FitB,colorlinks=false]{hyperref}
\hypersetup{
	pdftitle={Weak decay of halo nuclei},
	pdfauthor={Wael Elkamhawy},
	pdfview=FitB
	pdfstartview=FitB
	}
\usepackage[capitalize]{cleveref}
\usepackage{pgfplots}
\pgfplotsset{compat=newest}
\usepackage{simpler-wick}
\usepackage{color}
\usepackage{xcolor}
\usepackage{dsfont}
\usepackage{float}

\newcommand{\bea}{\begin{eqnarray}}
\newcommand{\eea}{\end{eqnarray}}
\newcommand{\beq}{\begin{equation}}
\newcommand{\eeq}{\end{equation}}
\newcommand{\be}{\begin{equation}}
\newcommand{\ee}{\end{equation}}
\newcommand{\bqa}{\begin{eqnarray}}
\newcommand{\eqa}{\end{eqnarray}}

\def\mqo2{{\!\!\!}}

\begin{document}
\title{Weak decay of halo nuclei}

\author{Wael Elkamhawy}
\email{elkamhawy@theorie.ikp.physik.tu-darmstadt.de}
\affiliation{Institut f\"ur Kernphysik, Technische Universit\"at Darmstadt, 64289 Darmstadt, Germany}

\author{Hans-Werner Hammer}
\email{Hans-Werner.Hammer@physik.tu-darmstadt.de}
\affiliation{Institut f\"ur Kernphysik, Technische Universit\"at Darmstadt, 64289 Darmstadt, Germany}
\affiliation{ExtreMe Matter Institute EMMI and Helmholtz Forschungsakademie Hessen f\"ur FAIR (HFHF), GSI Helmholtzzentrum
für Schwerionenforschung GmbH, 64291 Darmstadt, Germany}

\author{Lucas Platter}
\email{lplatter@utk.edu}
\affiliation{Department of Physics and Astronomy, University of Tennessee, Knoxville, TN 37996, USA}
\affiliation{Physics Division, Oak Ridge National Laboratory, Oak Ridge, TN 37831, USA}
\affiliation{Institut f\"ur Kernphysik, Technische Universit\"at Darmstadt, 64289 Darmstadt, Germany}
\affiliation{ExtreMe Matter Institute EMMI and Helmholtz Forschungsakademie Hessen f\"ur FAIR (HFHF), GSI Helmholtzzentrum
für Schwerionenforschung GmbH, 64291 Darmstadt, Germany}
 
\date{\today}

\begin{abstract}
  We investigate the weak decay of one-neutron halo nuclei into the
  proton-core continuum, i.e., beta-delayed proton emission from the halo
  nucleus using a cluster effective field theory for halo nuclei.
  On the one hand,
  we calculate the direct decay into the continuum. On the other hand, we
  consider the case of resonant final state interactions between the proton
  and the core. We present our formalism and discuss the
  application to the decay of $^{11}$Be in detail. Moreover, we
  compare to recent experimental results for the branching ratio and
  resonance parameters. As another example, we
  consider the case of $^{19}$C and predict the branching ratio
  for beta-delayed proton emission.
\end{abstract}

\smallskip
\maketitle

\newpage
\section{Introduction}
\label{sec:introduction}

Halo nuclei are exotic nuclei showing a pronounced cluster structure. They
consist of a tightly bound core
nucleus and a few weakly bound valence nucleons.
Halo nuclei thus exhibit a reduction in the number of
degrees of freedom since the many-body nucleus can
be described as an effective few-body system. This emergence of new
degrees of freedom is signaled by a separation of scales
that is also apparent in observables. In halo nuclei, the
scales $R_{\rm core}$ and $R_{\rm halo}$ that denote the length scales
of the core and halo, respectively, are clearly separated such that
the ratio $R_{\rm core}/R_{\rm halo}\ll 1$ can be used as an expansion
parameter. One can then construct an effective field theory (EFT) to
calculate halo observables in a systematic expansion in
$R_{\rm core}/R_{\rm halo}$ using only the new
{\it effective} degrees of freedom~\cite{Bertulani:2002sz,Bedaque:2003wa}.
This EFT for halo nuclei is known
as Halo EFT and has been applied successfully to many halo 
nuclei (see Refs.~\cite{Hammer:2017tjm,Hammer:2019poc,Hammer:2022lhx}
for recent reviews). This separation of
scales also leads to other universal features, such as 
the dependence of matrix elements on only a couple of few-body
parameters. For example, the one-neutron separation energy of an
$S$-wave one-neutron halo is directly related to the core-neutron scattering
length. Similar relations exist for other static properties, such as
the matter or charge radii, and reaction rates such as neutron
capture.

Another less obvious candidate for a halo physics-dominated observable
is the weak decay rate: One might expect that the decay rate of a
one-neutron halo nucleus is determined by the weakly bound neutron,
provided the core half-life is significantly larger than the neutron
half-life. However, such decays can be dominated by transitions into
deeply bound states of the daughter nucleus since the total decay
rate depends on the available phase space, which is larger when more
energy is released. Nonetheless, one can still identify the
channel in which a proton is emitted into the continuum
as originating from the decay of the halo neutron. For $^{11}$Be this
decay channel was first studied theoretically in a cluster model by Baye and
Tursonov~\cite{Baye:2010cj}. The first experimental results for this
rare decay mode were reported in Refs.~\cite{Borge:2012nz,Riisager:2014rma}.
Riisager {\it et al.}~\cite{Riisager:2014gia}
measured a surprisingly large branching ratio for this decay process,
which could only be understood in their analysis if the decay proceeds
through a new single-particle resonance in $^{11}$B. Their measured
branching ratio is also more than two orders of magnitude larger than
the cluster model prediction by Baye and Tursunov~\cite{Baye:2010cj}. This led
Pf\"utzner and Riisager~\cite{Pfutzner:2018ieu} to propose
$\beta$-delayed proton emission from $^{11}$Be as a possible avenue to detect
a dark matter decay mode, as suggested by Fornal and
Grinstein~\cite{Fornal:2018eol}.  Recently, Aayad {\it et
  al.}~\cite{Ayyad:2019kna} remeasured the branching ratio
for beta-delayed proton emission from $^{11}$Be and
reported a result similar to that given in Ref.~\cite{Riisager:2014gia}.
Subsequently, they also presented data supporting the existence of a
low-lying resonance in the $^{11}$B system~\cite{Ayyad:2022zqw}.
Their resonance energy is consistent with a near-threshold proton resonance
in $^{11}$B which was observed in the reaction
$^{10}\mathrm{Be}(d,n)^{11}\mathrm{B}^* \to\, ^{10}\mathrm{Be} + p$
in inverse kinematics~\cite{PhysRevLett.129.012502}.
Using Halo EFT, we have shown that the extracted resonance parameters
and decay rate by Ayaad {\it et al.} are
self-consistent~\cite{Elkamhawy:2019nxq}. The existence of such
a resonance in  $^{11}$B was also supported by a calculation in the shell
model embedded in the continuum and attributed to the coupling of
shell model states to the continuum \cite{Okolowicz:2019ifb}, although
the large branching ratio into $\beta^- p$  was cast into doubt by a
combined study of the $\beta^- p$ and $\beta^- \alpha$ decay
channels~\cite{Okolowicz:2021qgl}. Finally, recent calculations
in the no-core shell model with continuum~\cite{Atkinson:2022jfx}
and self-consistent Skyrme Hartree-Fock in the continuum also
found the near-threshold proton resonance in $^{11}$B~\cite{Nguyen:2022qzt}.
  
One important feature of Halo EFT is that observables are
parameterized in terms of a few observables. So although the rates for
$\beta$-delayed proton emission of different halo nuclei can be very
different, in first approximation they are governed by
the same analytical expressions. Here we
focus on these universal aspects of $\beta$-delayed proton
emission in halo nuclei. We give results for matrix elements in
terms of the few relevant parameters in this problem: the effective
range parameters in the initial and final state channels, the mass of
the halo nucleus, and the charge of the core nucleus.
Due to the large current interest, the main focus is on the application
to the beta-delayed proton emission from $^{11}$Be.
However, we also consider the case of $^{19}$C as further example. 

This manuscript is organized as follows. In
Sec.~\ref{sec:halo-effective-field}, we will set up the Halo EFT
theoretical framework, which includes the Lagrangian for the system
under consideration, the renormalization of the involved coupling
constants, and a derivation of the weak matrix elements that are
needed. Finally, we end with a summary.

\section{Halo effective field theory and weak decay}
\label{sec:halo-effective-field}
In the following, we use natural units with $\hbar=c=1$. 
The Halo EFT Lagrangian $ \mathcal{L}$ for a one-neutron-halo nucleus
as well as a low-lying resonance in the core-proton system up to
next-to-leading order can be written as
$\mathcal{L} = \mathcal{L}_0 + \mathcal{L}_{d}+ \ldots$,
where the $\dots$ indicate higher-order terms in the expansion
in $R_{\rm core}/R_{\rm halo}$.
$\mathcal{L}_0$ is the free Lagrangian of the core, neutron and proton
\begin{equation}
\begin{aligned}
  \label{eq:freeLagrangian}
  \mathcal{L}_0 = c^\dagger&\left(i\partial_t + \frac{\nabla^2}{2m_c}
  \right)c
  +   n^\dagger\left(i\partial_t + \frac{\nabla^2}{2m_n} \right)n \\
  &+   p^\dagger\left(i\partial_t + \frac{\nabla^2}{2m_p} \right)p ~,
\end{aligned}
\end{equation}
with $c$, $n$ and $p$ the core, neutron and proton fields, respectively.
In the case of $^{11}$Be, the 
masses of core, neutron and proton are denoted by 
$m_c=9327.548$~MeV, $m_n=939.565$~MeV and $m_p=938.272$~MeV.
For other $1n$ halos $m_c$ has to be adjusted accordingly.
The $S$-wave core-neutron as well as core-proton interaction
are described by  $\mathcal{L}_{d}$, which reads
\begin{equation}
\begin{aligned}
  \label{eq:Lsigma}
  \mathcal{L}_{d}= d_n^\dagger&\left[ \eta \left( i\partial_t +
      \frac{\nabla^2}{2M_{nc} } \right)+ \Delta \right] d_n \\
      &+d_p^\dagger\left[ \tilde{\eta} \left( i\partial_t +
      \frac{\nabla^2}{2M_{pc} } \right)+ \tilde{\Delta} \right] d_p \\
  &-g  \left[ c^\dagger n^\dagger d_n + \text{H.c.}  \right]
  -\tilde{g}  \left[ c^\dagger p^\dagger d_p + \text{H.c.}  \right],
\end{aligned}
\end{equation}
where $d_n$ and $d_p$ are dimer fields, with spin indices suppressed, that
represent the $J^P = 1/2^{+}$ ground state of the $1n$-halo nucleus 
and the $J^P = 1/2^{+}$ low-lying resonance in the core-proton system,
respectively, while $M_{nc}=m_n+m_c$ and $M_{pc}=m_p+m_c$.
This effective Lagrangian contains all terms required up to
next-to-leading order (NLO) in the strong interaction sector
in the power counting in $R_{\rm core}/R_{\rm halo}$.

The renormalization of the low-energy constants of the $S$-wave
$1n$-halo nucleus will be briefly summarized based on
Ref.~\cite{Hammer:2011ye}. We present the relevant results to define
our notation.  Due to the non-perturbative nature of the interaction,
we need to resum the self-energy diagrams to all orders. After
matching the low-energy constants for the $1n$-halo nucleus appearing in
Eq.~\eqref{eq:Lsigma} to the effective range expansion, we obtain the
full two-body $T$-matrix in the center-of-mass of the neutron-core
system
\begin{equation}
  \label{eq:t0}
  T_0(E)  
  =\frac{2\pi}{m_R} \left[\frac{1}{a_0} - r_0 m_R E - \sqrt{-2m_R E -i \epsilon} 
\right]^{-1}~,
\end{equation}
where 
$m_R$ is the reduced mass, and 
$a_0$, $r_0$ are the $S$-wave core-neutron scattering length and effective 
range, respectively.
This expression holds to NLO in the power counting in $R_{\rm core}/R_{\rm halo}$.
The corresponding
leading order (LO) result can be obtained by setting $r_0=0$.\footnote{
For convenience, we do not treat the range $r_0$ in strict perturbation
theory and keep it in the denominator of Eq.~(\ref{eq:t0}).}
The residue $Z$ of Eq.~(\ref{eq:t0}) at the bound state pole is 
required to calculate physical observables:
$Z=\textstyle{\frac{2\pi \gamma_0}{m_R^2}}/(1-r_0\gamma_0)$~,
with $\gamma_0 =(1-\sqrt{1-2r_0/a_0})/r_0\equiv
\sqrt{2m_R S_n}$ the binding momentum of the $S$-wave
halo state, and $S_n$ the one-neutron separation energy of the halo nucleus.

In order to investigate $\beta$-delayed proton emission from
a generic $1n$-halo nucleus, we include the weak interaction current allowing
transitions of a neutron into a proton, electron and antineutrino which
corresponds to the hadronic one-body current.
Moreover, we have to consider hadronic two-body currents that appear in the 
dimer formalism once the effective range is included.
The corresponding Lagrangian is given by
\begin{align}
  \label{eq:Lweak}
  \mathcal{L}_{\text{weak}} = -\frac{G_F}{\sqrt{2}} l_-^\mu \left(\left(J_\mu^+\right)^{\text{1b}} + \left(J_\mu^+\right)^{\text{2b}}\right)~,
\end{align}
where $l^{\mu}_- = \bar{u}_e \gamma^\mu (1-\gamma^5) v_{\bar{\nu}}$ 
and $\left(J_{\mu}^{+}\right)^{\text{1b}} = (V_\mu^1 - A_\mu^1)  + i (V_\mu^2 - A_\mu^2) $ 
denote the leptonic and hadronic one-body currents, respectively. 
Here the hadronic one-body current is decomposed into vector and axial-vector
contributions. At leading order, the contributions to this current are 
$ V_0^a  = N^\dagger \frac{\tau^a}{2}N$, 
$A_k^a = g_A N^\dagger \frac{\tau^a}{2}\sigma_k N$, 
where $g_A \simeq 1.27$ is the axial-vector coupling constant~\cite{Chang:2018uxx}. Terms with more derivatives and/or more fields
(many-body currents) will appear at higher orders. The first and
second term give the conventional Fermi and Gamow-Teller operators,
respectively.
Including resonant core-proton final state interactions, we have to take into account a two-body current with known coupling constants which arises from gauging the time derivative of the dimer fields appearing in Eq.~\eqref{eq:Lsigma}.
It is also decomposed into vector and axial-vector contributions and reads
\begin{align}
	\left(J_\mu^+\right)^{\text{2b}} = \begin{cases} -d_p^\dagger \, d_n \,\, &\mu=0~,\\ g_A \, 	d_p^\dagger \, \sigma_k \, d_n \,\, &\mu=k=1,2,3~.\end{cases}
	\label{eq:dimerterm}
\end{align}
In addition, there is also an unknown contribution usually denoted as
$L_{1A}$ that normally appears at the same order.  However, in the
case with Coulomb interaction, this piece is suppressed by
$\left(R_{\text{core}}/R_{\text{halo}}\right)^{1/2}$ compared to the
two-body current in Eq.~\eqref{eq:dimerterm}.\footnote{The scaling of
  $r_0^C \sim 1/k_C$ \cite{Ryberg:2015lea,Schmickler:2019ewl} leads to the suppression of the counterterm
  contribution $L_{1A}$.} Therefore, it contributes only at NNLO
allowing us to make predictions up to NLO. Note that our power
counting including resonant final state interactions implies a
suppression of $\left(R_{\text{core}}/R_{\text{halo}}\right)^{1/2}$
going from order to order instead of $R_{\text{core}}/R_{\text{halo}}$
as in the case without resonant final state interactions.

\subsection{Weak matrix element and decay rate}
The decay rate for the decay of the $1n$-halo nucleus into the final
particles given by core, proton, electron and antineutrino expressed
via the matrix element $\mathcal{M}$ reads
\begin{multline}
  \Gamma_p=\int \frac{d^3p_c}{(2\pi)^3} \int \frac{d^3p_p}{(2\pi)^3} \int \frac{d^3p_e}{(2\pi)^3(2E_e)} \int \frac{d^3p_{\bar{\nu}}}{(2\pi)^3(2E_{\bar{\nu}})} ~ \overline{|\mathcal{M}\left(\mathbf{p}_c,\mathbf{p}_p,\mathbf{p}_e,\mathbf{p}_{\bar{\nu}}\right)|^2}\\
  \times (2\pi)^4 ~ \delta\left(S_n- \Delta m + E_c + E_p + E_e + E_{\bar{\nu}}\right) ~ \delta^3\left(\mathbf{p}_c+\mathbf{p}_p+\mathbf{p}_e+\mathbf{p}_{\bar{\nu}}\right)\,,
\end{multline}
where $\Delta m = 1.29$ MeV is the mass difference between neutron and
proton. Here, $E_e = \sqrt{m_e^2 + \mathbf{p}_e^2}$ is the relativistic on-shell
energy of the electron with $m_e=0.511$~MeV denoting the electron
mass, $E_{\bar{\nu}}=|\mathbf{p}_{\bar{\nu}}|$ is the relativistic
energy of the antineutrino (assumed to be massless),
$E_c=\mathbf{p}_c^2/(2m_c)$ and
$E_p=\mathbf{p}_p^2/(2m_p)$ are the non-relativistic kinetic
energies of the core and proton, respectively.
Furthermore, $\mathbf{p}_i$
with $i\in\{c,p,e, \bar{\nu}\}$ is the momentum of the corresponding
particle.
$\overline{|\mathcal{M}|^2}\left(\mathbf{p}_c,\mathbf{p}_p,\mathbf{p}_e,\mathbf{p}_{\bar{\nu}}\right)$
denotes the
squared matrix element summed over final spins and averaged over
initial spins. Changing variables, we substitute the
coordinates $\mathbf{p}_c$ and $\mathbf{p}_p$ by the relative momentum
$\mathbf{p_{\text{rel}}}\equiv\mathbf{p}$ and the total momentum
$\mathbf{p_{\text{tot}}}$ of the core-proton system. Moreover, we neglect 
recoil effects in the energy-conserving delta-distribution and then use the
momentum-conserving delta-distribution to find
\begin{multline}
  \Gamma_p=\int \frac{d^3p}{(2\pi)^3} \int \frac{d^3p_e}{(2\pi)^3(2E_e)} \int \frac{d^3p_{\bar{\nu}}}{(2\pi)^3(2E_{\bar{\nu}})} ~ \overline{|\mathcal{M}\left(\mathbf{p},-(\mathbf{p}_e+\mathbf{p}_{\bar{\nu}}),\mathbf{p}_e,\mathbf{p}_{\bar{\nu}}\right)|^2}\\
  \times (2\pi) ~ \delta\left(S_n- \Delta m + \frac{\mathbf{p}^2}{2m_R} + E_e + E_{\bar{\nu}}\right)\,.
\end{multline}
We define the four-dimensional vector $\tilde{\sigma}_\mu$ consisting of $2\times2$-matrices for each component,
\begin{align}
  \tilde{\sigma}_\mu= \begin{cases} \mathds{1}_{2\times2} \,\, &\mu=0~,\\ -g_A \, \sigma_k \,\, &\mu=k=1,2,3~, \end{cases}
\end{align}
and divide the squared matrix element into a purely
leptonic and hadronic part. Separating the contribution of the
nucleon spin operators, $G_F^2/2 \Tr\left(\tilde{\sigma}_\mu
  \tilde{\sigma}_\nu^\dagger\right)$, from the hadronic part we obtain
\begin{align}
  \overline{|\mathcal{M}|^2}=\sum_{\mu,\nu} \sum_{\text{lept.spins}} |\mathcal{M}_{\text{lept.}}^{\mu,\nu}\left(\mathbf{p}_e,\mathbf{p}_{\bar{\nu}}\right)|^2 ~ \left(\frac{G_F^2}{2}\Tr\left(\tilde{\sigma}_\mu \tilde{\sigma}_\nu^\dagger\right)\right) \frac{1}{2J+1}|\mathcal{A}\left(\mathbf{p},-\left(\mathbf{p}_e+\mathbf{p}_{\bar{\nu}}\right)\right)|^2\,,
\end{align}
where $J=1/2$ denotes the total spin of the $S$-wave halo nucleus.
Further,
$|\mathcal{M}_{\text{lept.}}^{\mu,\nu}\left(\mathbf{p}_e,\mathbf{p}_{\bar{\nu}}\right)|^2$
and
$|\mathcal{A}\left(\mathbf{p},-\left(\mathbf{p}_e+\mathbf{p}_{\bar{\nu}}\right)\right)|^2$
denote the leptonic and hadronic part, respectively. Note that we
have already set
$\mathbf{p}_{\text{tot}}=-\left(\mathbf{p}_e+\mathbf{p}_{\bar{\nu}}\right)$. The
leptonic part reads
\begin{align}
   \sum_{\text{lept.spins}}|\mathcal{M}_{\text{lept.}}^{\mu,\nu}\left(\mathbf{p}_e,\mathbf{p}_{\bar{\nu}}\right)|^2&=\sum_{\text{lept.spins}} \left(\bar{u}_e \gamma^\mu (1-\gamma^5) v_{\bar{\nu}}\right) \left(\bar{u}_e \gamma^\mu (1-\gamma^5) v_{\bar{\nu}}\right)^\dagger\\
   &=8\left(P_e^\mu P_{\bar{\nu}}^\nu+P_e^\nu P_{\bar{\nu}}^\mu-g^{\mu\nu}\left(P_e \cdot P_{\bar{\nu}}\right)\right)\,,
   \label{eq:leptonicpart}
\end{align}
where $P_\alpha$ with $\alpha \in \{e,\bar{\nu}\}$ is the four-momentum of the corresponding particle as indicated by its subscript.
Evaluating the trace over spin operators leads to
\begin{align}
  \label{eq:weakpart}
  \frac{G_F^2}{2}\Tr\left(\tilde{\sigma}_\mu \tilde{\sigma}_\nu^\dagger\right)=\begin{cases} G_F^2 \,\, &\mu=\nu=0\,,\\ G_F^2\,g_A^2 \delta_{kl} \,\, &\mu=k\,, \,\, \nu=l \,\,\, \text{with} \,\,\, k,l\in\{1,2,3\}\,,\\ 0 \,\, &\text{otherwise.} \end{cases}
\end{align}
From \cref{eq:weakpart}, we conclude that we have a non-vanishing contribution
either for $\mu=\nu=0$ or for \mbox{$\mu=k$} and $\nu=l$. The first
case implies no spin-flip during the decay of the neutron into proton
while the latter implies a spin-flip during that transition. The first
contribution corresponds to the Fermi and the latter to the
Gamow-Teller transition. Note that there is no interference
term. Using \cref{eq:leptonicpart,eq:weakpart}, we find
\begin{align}
  \label{eq:matrixelement}
  \overline{|\mathcal{M}|^2}=4G_F^2\Big(\sqrt{\mathbf{p}_e^2+m_e^2} ~|\mathbf{p}_{\bar{\nu}}|(1+3g_A^2)+\mathbf{p}_e \cdot \mathbf{p}_{\bar{\nu}}(1-g_A^2)\Big) ~ |\mathcal{A}\left(\mathbf{p},-\left(\mathbf{p}_e+\mathbf{p}_{\bar{\nu}}\right)\right)|^2\,.
\end{align}
In order to take into account electromagnetic interactions of the
emitted electron with the remaining charged particles, we multiply
\cref{eq:matrixelement} with the Sommerfeld factor of the electron
given by
\begin{align}
  \label{eq:Sommerf}
  C^2(\eta_e)= \frac{2\pi\eta_e}{(e^{2\pi\eta_e} - 1)}~,
\end{align}
where $\eta_e = \alpha Z Z_e E_e/|\mathbf{p}_e|$
with $\alpha =1/137$ the fine structure 
constant. We use $Z=Z_p$ in order to ensure that we reproduce the 
free neutron decay width in the limit of a vanishing one-neutron 
separation energy of the $1n$-halo nucleus. This means that the electron is
only interacting with the outgoing proton. We assume this to be a good
approximation since the core is far away from the decaying valence neutron
due to the small one-neutron separation energy. The error introduced by
this approximation is of higher order (see Ref.~\cite{Elkamhawy:2019nxq}
for an explicit estimate).
Therefore, \cref{eq:matrixelement} becomes
\begin{align}
  \label{eq:matrixelement2}
  \overline{|\mathcal{M}|^2}=4G_F^2\Big(E_e ~E_{\bar{\nu}}(1+3g_A^2)+\mathbf{p}_e \cdot \mathbf{p}_{\bar{\nu}}(1-g_A^2)\Big) ~ |\mathcal{A}\left(\mathbf{p},-\left(\mathbf{p}_e+\mathbf{p}_{\bar{\nu}}\right)\right)|^2 ~  C^2(\eta_e)\,.
\end{align}
Substituting energies for momenta, integrating out the energy
conserving $\delta$-function and adjusting the integration momenta
accordingly, we then obtain for the decay rate\footnote{A detailed derivation of the rate equation below is given in App.~\ref{app:rate}.}
\begin{multline}
\label{eq:partial-rate}
  \Gamma_p=\frac{G_F^2}{16\pi^5} \int_0^{\Delta m -S_n-m_e} dE \int_{m_e}^{\Delta m -S_n -E} dE_e \int_{-1}^1 d\cos\theta \int_{-1}^1 d\cos\theta_{\bar{\nu}} ~ m_R \sqrt{2m_R E}\\
  \times E_e ~ \sqrt{E_e^2-m_e^2} ~ \left(\Delta m - S_n -E -E_e\right)^2 \Big((1+3g_A^2)+\beta_e \cos\theta_{\bar{\nu}} (1-g_A^2)\Big)\\
  \times |\mathcal{A}\left(E,E_e,\cos\theta,\cos\theta_{\bar{\nu}}\right)|^2 ~  C^2(\eta_e)\,,
\end{multline}
where $\pmb{\beta}_e=\mathbf{p}_e/E_e$ and
$\pmb{\beta}_{\bar{\nu}}=\mathbf{p}_{\bar{\nu}}/E_{\bar{\nu}}$.
The differential decay rate reads
\begin{multline}
  \frac{d\Gamma_p}{dE}=\frac{G_F^2}{16\pi^5} ~ m_R \sqrt{2m_R E} \int_{m_e}^{\Delta m -S_n -E} dE_e \int_{-1}^1 d\cos\theta \int_{-1}^1 d\cos\theta_{\bar{\nu}}\\
  \times E_e ~ \sqrt{E_e^2-m_e^2} ~ \left(\Delta m - S_n -E -E_e\right)^2 \Big((1+3g_A^2)+\beta_e \cos\theta_{\bar{\nu}} (1-g_A^2)\Big)\\
  \times |\mathcal{A}\left(E,E_e,\cos\theta,\cos\theta_{\bar{\nu}}\right)|^2 ~  C^2(\eta_e)
\end{multline}
with
\begin{align}
  0<E<\Delta m -S_n -m_e\,.
\end{align}
From the partial decay rate we obtain the branching ratio via
\begin{align}
  \label{eq:bp}
  b_p=\frac{\Gamma_p}{\Gamma}\,,
\end{align}
where $\Gamma=\ln(2)/T_{1/2}$ is the full decay rate and $T_{1/2}$ the half-life of the halo nucleus.

\subsection{Hadronic Amplitude without final state interactions}
\begin{figure}[t]
  \centering
  \includegraphics[width=0.2\textwidth]{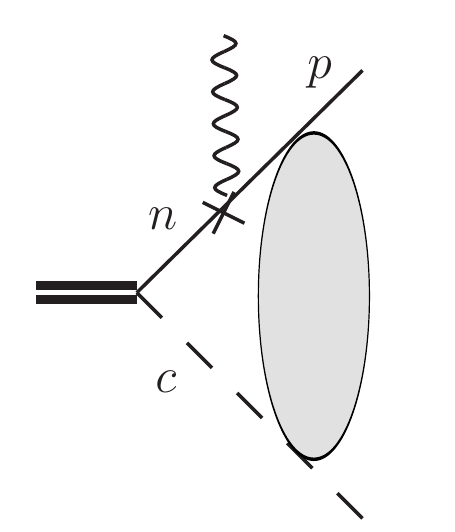}
  \caption{\label{fig:Feynman1} Feynman diagram for
      the weak decay of a one-neutron halo nucleus
      into the corresponding core and a proton with Coulomb final
      state interactions only. The shaded ellipse denotes the Coulomb ladder diagrams where multiple (including zero) photon exchanges contribute.}
\end{figure}

We first consider the hadronic amplitude without final state interactions.
It describes the coupling of
the electroweak current to the nucleus. For simplicity, we use
the momentum variables $\mathbf{p}$, $\mathbf{p}_e$ and
$\mathbf{p}_{\bar{\nu}}$ and later apply
the constraints from energy and momentum conservation.
The corresponding diagram is illustrated in Fig.~\ref{fig:Feynman1}. In Halo EFT, the amplitude is derived from the Feynman rules according to the Lagrangians given in \cref{eq:freeLagrangian,eq:Lsigma,eq:Lweak}. The pure hadronic part of Fig.~\ref{fig:Feynman1} is given by the loop with three propagators, the Coulomb ladder diagrams, the breakup and the wave function renormalization constant of the halo nucleus. After performing the energy integration of the loop, we are left with two propagators. One propagator together with the Coulomb ladder diagrams gives the outgoing Coulomb wave function $\left(\tilde{\chi}_{\mathbf{p}}^-(\mathbf{r})\right)^*$. The other propagator together with the breakup and wave function renormalization constant of the halo nucleus gives the bound state wave function of the $1n$-halo nucleus $\psi(\mathbf{r})$. The pure hadronic amplitude without final state interactions then reads
\begin{align}
  \mathcal{A}\left(\mathbf{p},\mathbf{p}_e,\mathbf{p}_{\bar{\nu}}\right)=-i\int d^3r \left(\tilde{\chi}_{\mathbf{p}}^-(\mathbf{r})\right)^* e^{i(1-y)(\mathbf{p}_e+\mathbf{p}_{\bar{\nu}})\cdot \mathbf{r}} \psi(\mathbf{r})\,,
\end{align}
where the expontential function results from the recoil due to the leptons.
In general, the Coulomb wave function carries all possible
angular momenta $l$ and can be written as 
\begin{align}
  \left(\tilde{\chi}_{\mathbf{p}}^-(\mathbf{r})\right)^*=\sum_{l=0}^{\infty} \left(\tilde{\chi}_{\mathbf{p}}^{l~-}(\mathbf{r})\right)^*=\sum_{l=0}^{\infty} \tilde{\chi}_{-\mathbf{p}}^{l~+}(\mathbf{r})=\sum_{l=0}^{\infty} (-1)^l ~ \tilde{\chi}_{\mathbf{p}}^{l~+}(\mathbf{r})
\end{align}
with (see Ref.~\cite{Higa:2016igc})
\begin{align}
  \tilde{\chi}_{\mathbf{p}}^{l~+}(\mathbf{r})=4\pi i^l e^{i\sigma_l} \frac{F_l(\eta_{|\mathbf{p}|},|\mathbf{p}|r)}{|\mathbf{p}|r}\sum_{m=-l}^{l}Y_{lm}^*(\mathbf{e}_{\mathbf{p}}) Y_{lm}(\mathbf{e}_{\mathbf{r}})
\end{align}
and
\begin{align}
  F_l(\eta_{|\mathbf{p}|})&=C_l(\eta_{|\mathbf{p}|})2^{-l-1}(-i)^{l+1}M_{i\eta_{|\mathbf{p}|},l+1/2}(2i|\mathbf{p}|r)\,,\\
  C_l(\eta_{|\mathbf{p}|})&=\frac{2^l e^{-\pi \eta_{|\mathbf{p}|}/2} |\Gamma(l+1+i\eta_{|\mathbf{p}|})|}{\Gamma(2l+2)}\,,
\end{align}
where $M_{k,\mu}(z)$ is the conventionally defined Whittaker function.
The Sommerfeld parameter is
$\eta_{|\mathbf{p}|}=\alpha Z_p Z_c m_R/|\mathbf{p}|=k_C/|\mathbf{p}|$
while $\sigma_l=\arg(\Gamma(l+1+i\eta_{|\mathbf{p}|}))$ is the Coulomb
phase shift. Moreover, the bound state wave function
of the neutron-core system is given by
\begin{align}
  \psi(\mathbf{r}) = \sqrt{2\gamma_0} \, \frac{e^{-\gamma_0 r}}{r} Y_{00}\,.
\end{align}
We use the plane-wave expansion 
\begin{align}
  e^{i(1-y)(\mathbf{p}_e+\mathbf{p}_{\bar{\nu}})\cdot \mathbf{r}}=4\pi \sum_{l=0}^{\infty}\sum_{m=-l}^{l} i^l j_l\left((1-y)|\mathbf{p}_e+\mathbf{p}_{\bar{\nu}}| r\right) Y_{lm}(\mathbf{e}_{\mathbf{p}_e+\mathbf{p}_{\bar{\nu}}}) Y_{lm}^*(\mathbf{e}_{\mathbf{r}})\,,
\end{align}
where $j_l(x)$ is a spherical Bessel function. In the low-energy limit, we approximate $j_l(x)\approx{(x)^l/(2l+1)!!}$. The hadronic amplitude then reads
\begin{equation}
  \begin{aligned}
  \mathcal{A}\left(\mathbf{p},\mathbf{p}_e,\mathbf{p}_{\bar{\nu}}\right)
  =&-i\sqrt{2\gamma_0}\sqrt{4\pi} \int dr \sum_{l=0}^{\infty}\sum_{m=-l}^{l} r e^{-\gamma_0 r}  e^{i\sigma_l} \frac{F_l(\eta_{|\mathbf{p}|},|\mathbf{p}|r)}{|\mathbf{p}|r}\\
  &\times 4\pi \frac{\left((1-y)|\mathbf{p}_e+\mathbf{p}_{\bar{\nu}}| r\right)^{l}}{(2l+1)!!}Y_{lm}(\mathbf{e}_{\mathbf{p}_e+\mathbf{p}_{\bar{\nu}}})Y_{lm}^*(\mathbf{e}_{\mathbf{p}})
  \end{aligned}
\end{equation}
The dominant contribution of the hadronic amplitude results from the
$l=0$ transition, meaning no angular momentum between the lepton
momentum $(\mathbf{p}_e+\mathbf{p}_{\bar{\nu}})$ and the outgoing
relative momentum $\mathbf{p}$. Taking only this contribution into
account leads to
\begin{equation}
  \begin{aligned}
  \mathcal{A}\left(\mathbf{p},\mathbf{p}_e,\mathbf{p}_{\bar{\nu}}\right)=&-i\sqrt{8\pi\gamma_0} e^{i\sigma_0} \int dr ~ r e^{-\gamma_0 r} \frac{F_0(\eta_{|\mathbf{p}|},|\mathbf{p}|r)}{|\mathbf{p}|r}\\
  =& -i\sqrt{8\pi\gamma_0} e^{i\sigma_0} C_0(\eta_{|\mathbf{p}|}) \frac{e^{2\eta_{|\mathbf{p}|}\arctan(|\mathbf{p}|/\gamma_0)}}{\gamma_0^2+|\mathbf{p}|^2}\,,
  \end{aligned}
\end{equation}
Higher $l$-transitions correspond to so-called forbidden decays, e.g.,
the $l=1$ transition is the first forbidden decay.

\begin{figure}[t]
  \centering
  \includegraphics[width=0.7\textwidth]{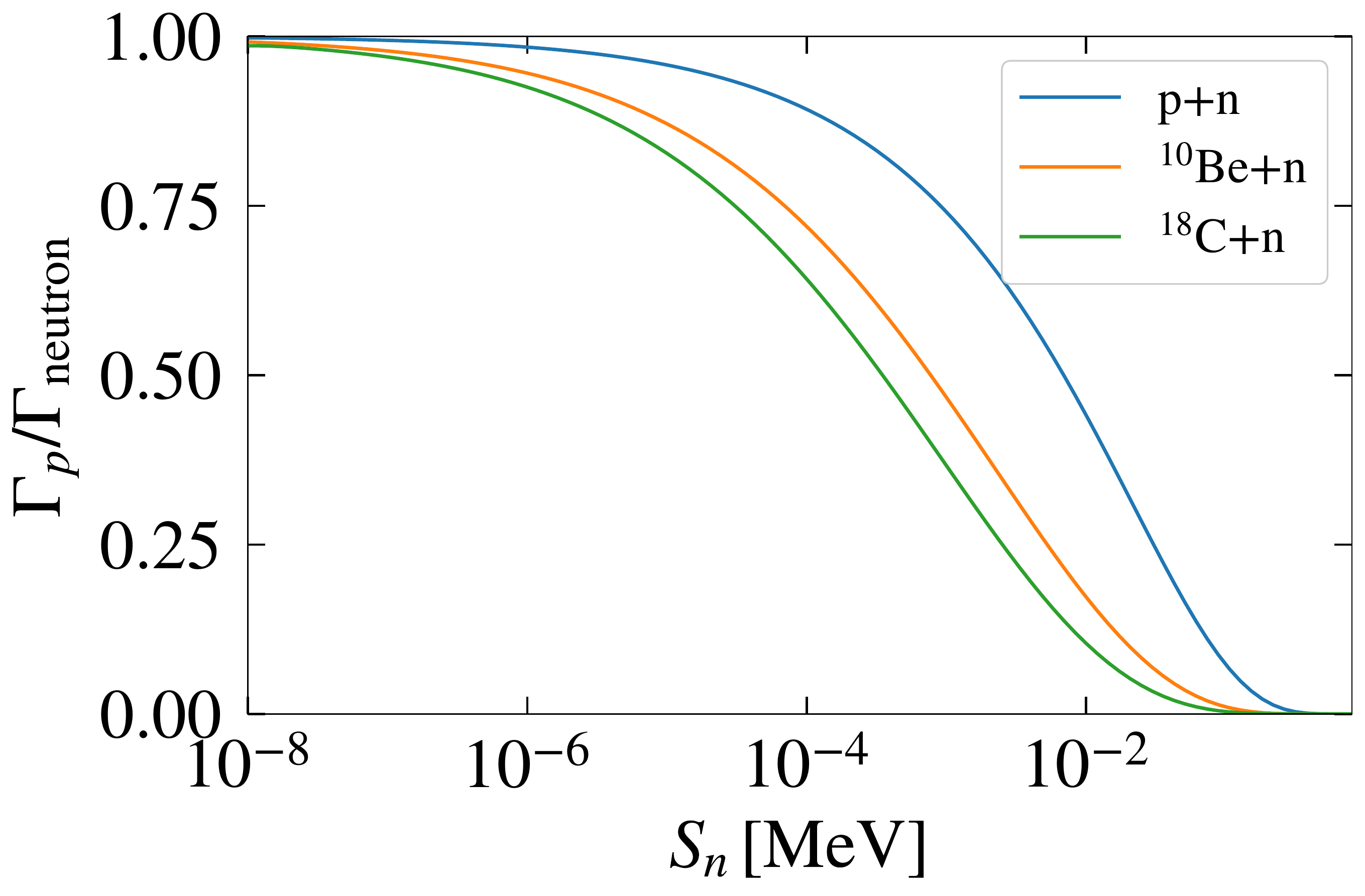}
  \caption{Partial decay rate without final state interactions at LO
    relative to the free neutron decay width
    $\Gamma_p/\Gamma_{\text{neutron}}$ as a function of the one-neutron
    separation energy $S_n$ for the systems p+n, $^{10}$Be+n and
    $^{18}$C+n.}
  \label{fig:directdecay}
\end{figure}

\section{Universal results without final state interactions}
\label{sec:results-without-final}
The rate for $\beta$-delayed proton emission
strongly depends on the form of the final state interactions. In
the absence of strong final state interactions, this rate is determined predominately by the one-neutron separation energy $S_n$ and the electric charge 
of the core. We first focus on the latter case.

In Fig.~\ref{fig:directdecay}, we show the partial decay rate calculated
with Eq.~\eqref{eq:partial-rate} in units of the neutron decay
rate $\Gamma_{\rm neutron}$ and in absence of strong final state interactions for systems with different core nuclei as a function of $S_n$. On the one hand, the results show that the decay rate becomes equal to the free neutron decay rate in the limit of zero one-neutron separation energy for all different systems. In this limit, the halo neutron is not influenced by the core at all and therefore it is expected to give the free neutron decay rate independently of the core properties. On the other hand, the decay rate approaches zero as the one-neutron separation energy increases up to the maximum value given by $S_n^{\text{max}}=m_n - m_p - m_e\approx782$~keV. This is also expected as the phase space is reduced for increasing $S_n$. 
In between these limits, the results show that the decay rate is further reduced for systems with core nuclei of larger sizes with respect to the mass and electric charge. It turns out that the mass dependency is negligible and therefore the reduction of the rate is dominantly given by the increase of the electric charge of the core. This is due to the Coulomb repulsion between the proton and core which is parameterized through the Sommerfeld factor $C_0(\eta_{|\mathbf{p}|})$. It leads to a reduction of the differential decay rate $d\Gamma_p/dE$ especially for low relative energies $E$ of the charged particles. Hence, the behaviour for different systems universally depends on the charge of the core.
Next we focus on the case with final state interactions.

\section{Results with final state interactions for Beryllium-11}
\label{sec:results-with-final}
We start by briefly reviewing the current status on
$\beta$-delayed proton emission from $^{11}$Be. The first
measurement of this decay channel was first reported in
Ref.~\cite{Borge:2012nz}. However, the collaboration withdrew its
initial claim of a very large branching ratio \cite{Riisager:2014gia}
and instead provided an upper limit for the branching ratio of
$b_p < 2.2 \times 10^{-6}$ \cite{Riisager:2020glj}.
In 2019, Ayyad {\it et al.}~\cite{Ayyad:2019kna} reported a measurement of
this branching ratio finding $b_p = (1.3 \pm 0.3) \times 10^{-5}$ and resonance
energy of $E_R = (196 \pm 20)$~keV with a width of $\Gamma_R = (12 \pm 5) $~keV.
Recently, Ayyad {\it et al.}~\cite{Ayyad:2022zqw} remeasured the
properties of the low-lying $^{11}$B resonance using $^{10}$Be-proton
scattering. Using an R-matrix analysis, they found
$E_{R} = \left(171\pm 20\right)$~keV and
$\Gamma_R = \left(4.5\pm1.1\right)$~keV for resonance energy and
width, respectively.
Furthermore, Lopez-Saavedra {\it et al.} used the reaction process
$^{10}$Be(d,p)$^{11}$B $\rightarrow$ $^{10}$B$+p$ to obtain the
estimate $E_R = (211\pm 40)$~keV for the resonance
energy~\cite{Lopez-Saavedra:2022vxh}.

A number of theoretical calculations has recently tried to address the
question of the possible existence of a resonance in the
proton-$^{10}$Be system. All
approaches that we are aware of found indeed such evidence. A
calculation in the shell model embedded in the
continuum \cite{Okolowicz:2019ifb} employed a phenomenological
shell model interaction and found evidence for a $1/2^+$ resonance at
approximately $E_R \approx 142$~keV.  Recently, Atkinson
{\it et al.}~\cite{Atkinson:2022jfx} used
the no-core shell model in the continuum to calculate the
phaseshifts for $^{10}$Be-proton scattering in channels of total
angular momentum $J$ and parity $\Pi$ and well-defined total isospin
$T$. Using a chiral effective interaction, the authors adjusted
their {\it ab initio} results to reproduce the $^{11}$B resonance position
and calculated the branching ratio of the decay into the continuum.
Their calculation suggests a definite isospin $T=1/2$ for the $^{11}$B
resonance.

\begin{figure}[t]
  \centering
  \includegraphics[width=0.7\textwidth]{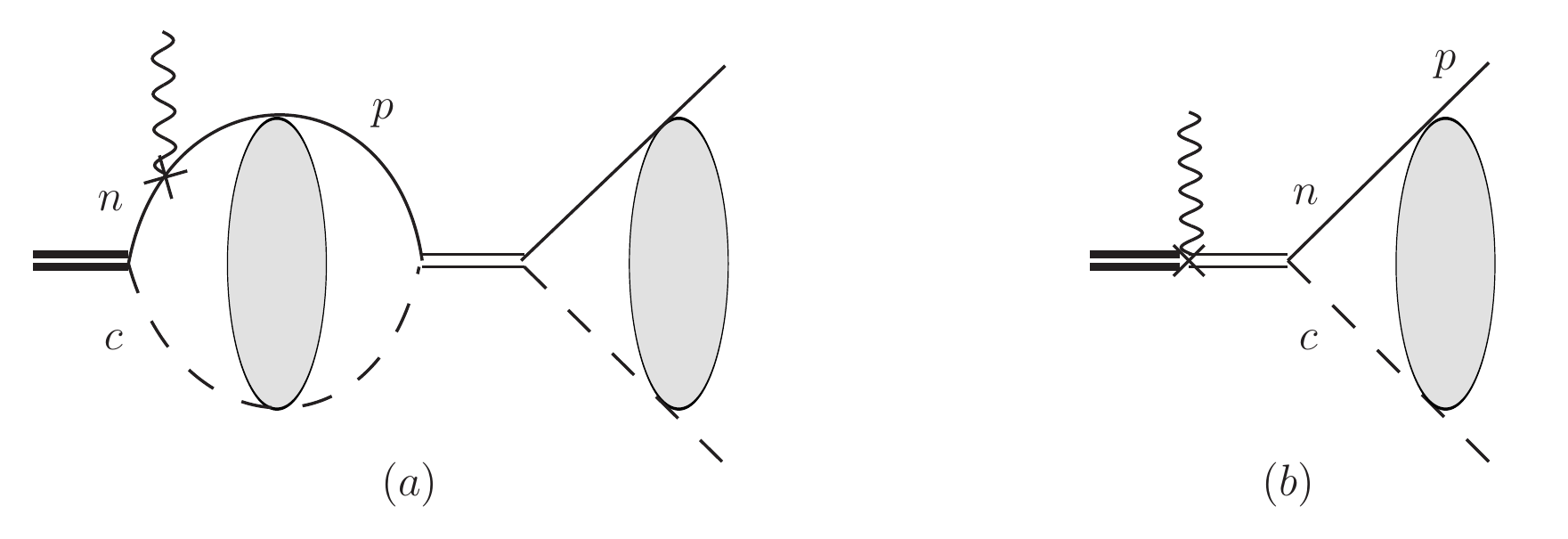}
  \caption{\label{fig:Feynman} Feynman diagrams for
      the weak decay of a one-neutron halo nucleus
      into the corresponding core and a proton with strong final
      state interactions. The thin double
      line in the middle denotes the dressed $^{10}$Be$-p$
      propagator. The shaded ellipse denotes the Coulomb ladder diagrams where multiple (including zero) photon exchanges contribute.}
\end{figure}
As laid out above, Halo EFT can not predict by itself whether the
resonance exists but instead provides a consistency check between
different observables in the resonance region.
In Ref.~\cite{Elkamhawy:2019nxq}, we performed an analysis of the
impact of a resonance on the branching ratio. In this analysis, it was assumed
that the resonance has no definite isospin. We briefly summarize our results
for this scenario.
Using the central value and errors of the resonance energy from the recent
publication~\cite{Ayyad:2022zqw}, the branching ratio and the resonance
width were determined to be \mbox{$b_p = \left(1.2_{-0.6}^{+1.1}\text{(exp.)}_{-0.2}^{+0.9}\text{(theo.)}\right)\times 10^{-5}$} and $\Gamma_R = \left(5.0^{+3.0}_{-2.1}\text{(exp.)}^{+3.1}_{-1.1}\text{(theo.)}\right)\text{~keV}$.
Based on the value and errors for the resonance energy published
previously~\cite{Ayyad:2019kna}, we obtain $b_p = \left(4.9_{-2.9}^{+5.6}\text{(exp.)}_{-0.8}^{+4.0}\text{(theo.)}\right)\times 10^{-6}$ and $\Gamma_R = \left(9.0^{+4.8}_{-3.3}\text{(exp.)}^{+5.3}_{-2.2}\text{(theo.)}\right)\text{~keV}$.
These results are consistent within the combined theoretical and
experimental uncertainties.

\begin{figure}[t]
  \centering
  \includegraphics[width=0.7\textwidth]{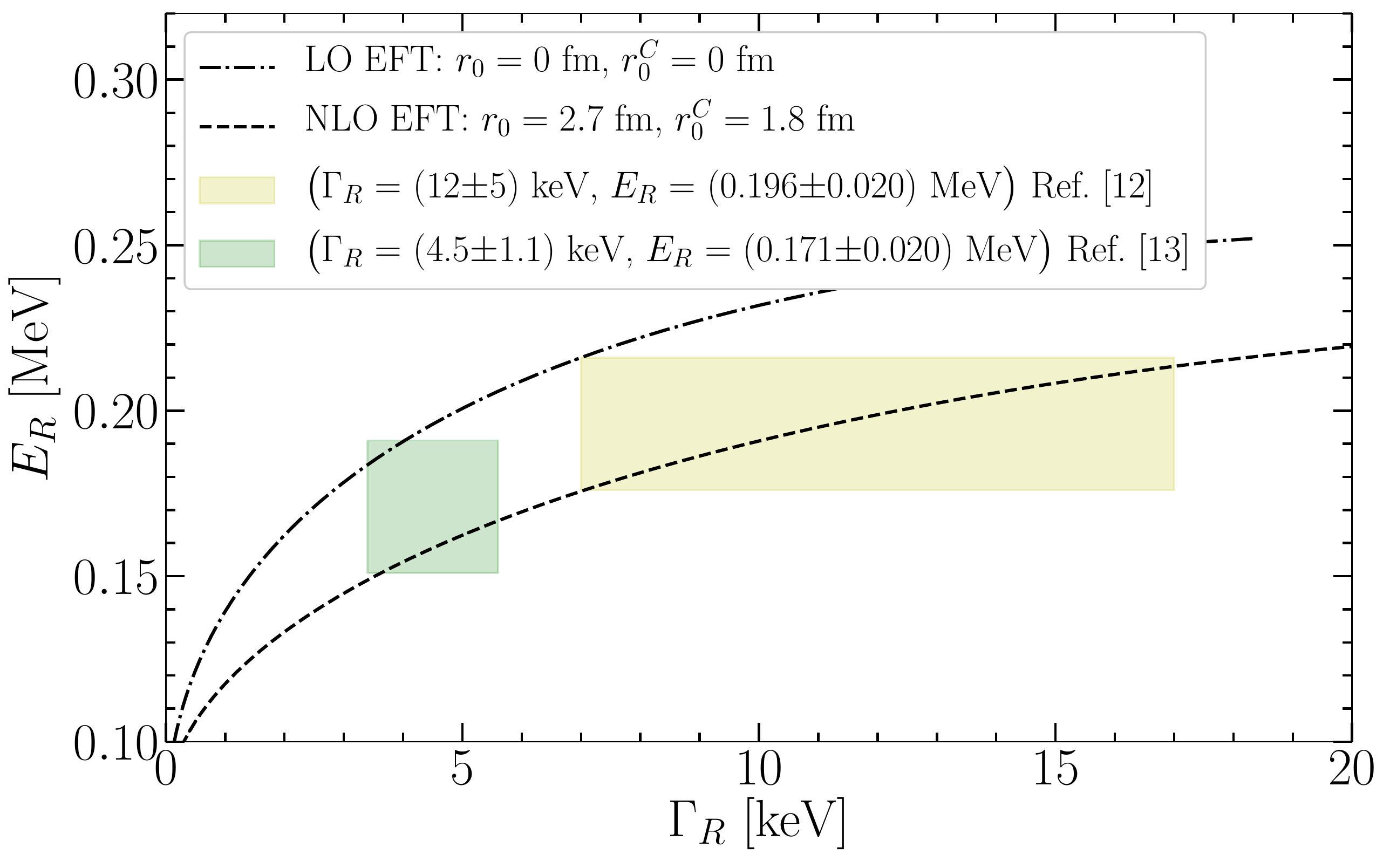}
  \caption{Possible resonance parameter combinations obeying the sum
    rule obtained in the Halo EFT approach with isospin projection.
    The dash-dotted line shows the combinations for $r_0=0$~fm at LO
    corresponding to $r_0^C=0$~fm while the dashed line shows the
    combinations for $r_0=2.7$~fm at NLO corresponding to
    $r_0^C=1.8$~fm. The green and yellow band show the resonance
    parameters given in Refs.~\cite{Ayyad:2019kna,Ayyad:2022zqw}.  }
  \label{fig:parameter_combinations_definite_isospin}
\end{figure}
Next we analyze the scenario of a $^{11}$B resonance with definite isospin
of $T=1/2$. The additional diagrams due to the resonance are depicted in Fig.~\ref{fig:Feynman}.
Starting with the Lagrangian given in Eqs.~\eqref{eq:freeLagrangian} and
\eqref{eq:Lsigma}, we project the $^{10}$Be$-p$ interaction on $T=1/2$.
Therefore, it is ensured that only the $T=1/2$ channel is resonant.
As a consequence, the isospin changes during the decay. This implies that
the transition is a pure Gamow-Teller transition. Moreover, this projection
impacts the beta-strength sum rule that counts the number of weak charges
that can decay in the initial state. The Gamow-Teller strength is related to
the comparative half-live of the decay, the so-called $ft$ value given by
\begin{align}
  ft = \frac{B}{g_A^2 B_{\text{GT}}}\,,
\end{align}
where $B=2\pi^3\ln2/(m_e^5G_F^2)$ is the $\beta$-decay constant. In this
paper, we use the value $B=6144.2$~s~\cite{Pfutzner:2011ju,Hardy:2004id}.
The inverse $ft$ value is directly related to the transition matrix
element $\mathcal{M}$ of $^{11}$Be into $^{10}\text{Be}+p$,
\begin{equation}
\begin{aligned}
  \frac{1}{ft}&=\frac{1}{B} ~ \overline{\left|\mathcal{M}\right|^2}\\
  &=\frac{1}{B} \frac{(1+3g_A^2)}{2\pi^2} \int dE \,  m_R \sqrt{2 m_R E} ~ \overline{|\mathcal{A}(\mathbf{p})|^2}\,.
\end{aligned}
\end{equation}
For a transition into the continuum, the differential Gamow-Teller
beta strength then reads
\begin{align}
	\frac{dB_{\text{GT}}}{dE} &= \frac{(1+3g_A^2)}{g_A^2} \frac{1}{2\pi^2} ~ m_R \sqrt{2 m_R E} ~~ 
	\overline{|\mathcal{A}(\mathbf{p})|^2} \,.
\end{align}
Integration over the whole continuum gives the beta-strength sum rule
that we require to be fulfilled at each order within our EFT power counting.
We note that a resonance with no definite isospin in the halo picture leads
to the sum rule $B^{\text{max}}_{\text{F}}=1$ and $B^{\text{max}}_{\text{GT}}=3$
(see Ref.~\cite{Elkamhawy:2019nxq}) accounting for the halo neutron that
can decay in the initial channel. However, after projecting on a resonance
with $T=1/2$, we do not fully count the weak charge that can decay
in the initial channel and therefore expect to have $B^{\text{max}}_{\text{GT}}<3$
for the beta-strength sum rule. At LO where the full non-perturbative
solution for a zero-range interaction is used in the incoming as well as
outgoing channels, we indeed find the sum rule satisfying
$B^{\text{max}}_{\text{GT}}<3$. When integrating over the available $Q$-window,
we therefore expect $B_{\text{GT}}<B^{\text{max}}_{\text{GT}}<3$. At NLO where
range corrections are included, the sum rule $B^{\text{max}}_{\text{GT}}$ derived
at LO puts strong constraints on the ranges in the incoming and outgoing
channels such that only certain combinations are allowed.
In Fig.~\ref{fig:parameter_combinations_definite_isospin}, we show the possible
resonance parameter combinations that fulfill the beta-strength sum rule. 
The dash-dotted line gives the result at LO where the
effective range in the incoming channel as well as the
Coulomb-modified effective range in the outgoing channel are set to
zero. At NLO, we use $r_0 = 2.7$~fm as determined in Ref.~\cite{Hammer:2011ye} 
from the measured B(E1) strength for Coulomb dissociation of $^{11}$Be. The
one-neutron separation energy as well as the effective range of
$^{11}$Be determine the Coulomb-modified effective range in the
outgoing channel to be $r_0^C= 1.8$~fm. The sum rule is then satisfied
to very good approximation for a wide range of Coulomb-modified
scattering lengths in the outgoing channel. We find that
both sets of resonance parameters from
Refs.~\cite{Ayyad:2019kna,Ayyad:2022zqw} are consistent with the constraints
on $E_R$ and $\Gamma_R$ provided by our Halo EFT calculation, although there
is some tension between the two measurements themselves.

\begin{figure}[t]
  \centering
  \includegraphics[width=0.7\textwidth]{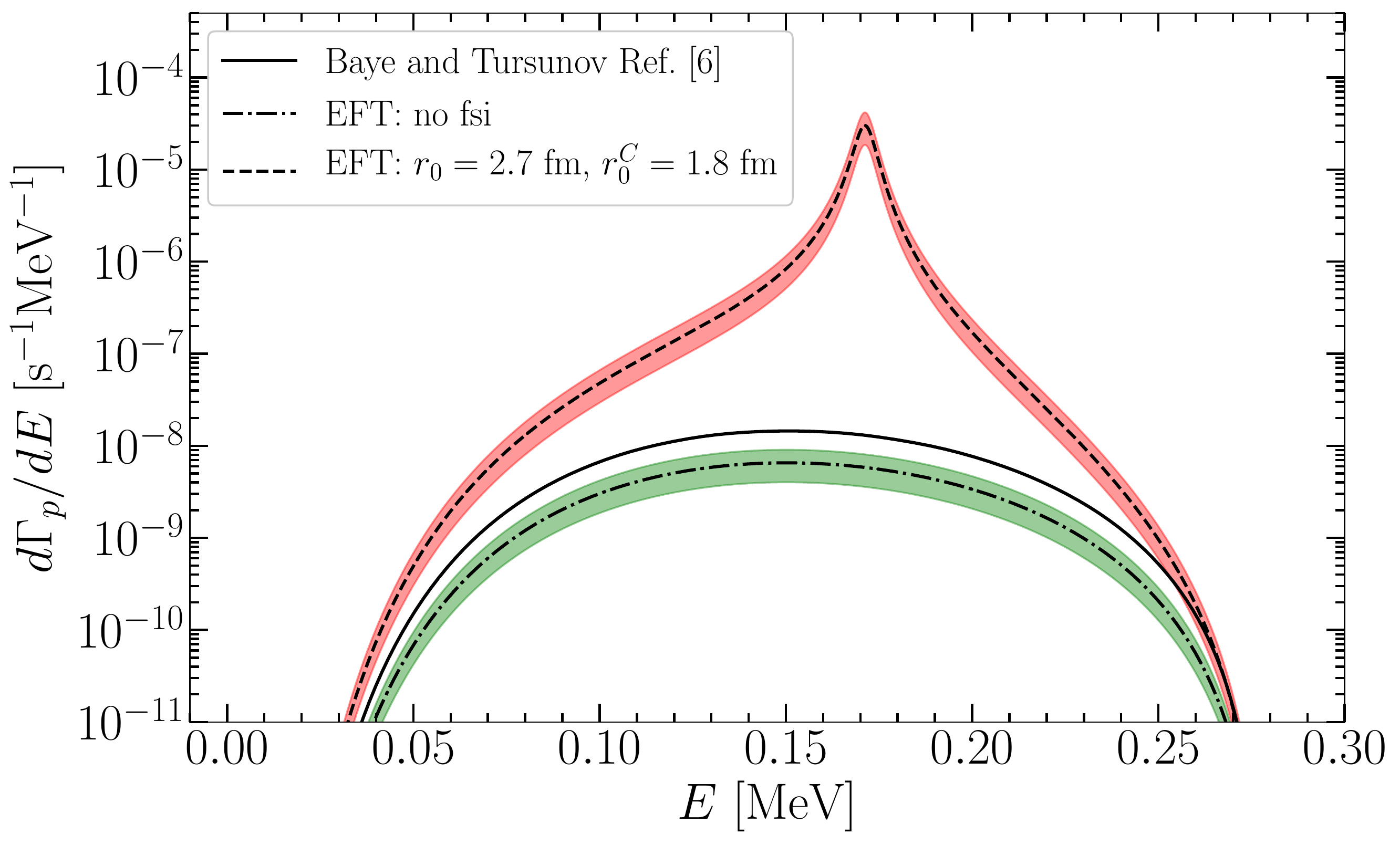}
  \caption{Differential decay rate $d\Gamma_p/dE$ for $\beta$-delayed
    proton emission from $^{11}$Be as a function of the final-state
    particle energy $E$.  The dash-dotted line shows our EFT result
    without resonant final state interactions while the solid line
    gives the result obtained by Baye and Tursunov \cite{Baye:2010cj}.
    The dashed line shows the EFT result including a resonance at
    $E_R=0.171$~MeV in the outgoing channel at NLO. The colored bands
    give the EFT uncertainty.}
  \label{fig:diffGamma_11Be_2}
\end{figure}

The effect of the resonance on the decay rate can be visualized by
looking at the differential decay rate $d\Gamma_p/dE$ as shown in
Fig.~\ref{fig:diffGamma_11Be_2}. The dash-dotted line shows the EFT
result in the absence of any final state interactions. The dashed line
shows the differential decay rate with a resonance at $E_R = 0.171$~MeV \cite{Ayyad:2022zqw} and effective ranges in initial and final state fixed as discussed
above. The rapid fall off of the differential decay rate at energies
below $\sim 0.08$~ MeV and above $\sim 0.21$~MeV, and the noticeable enhancement
of the decay rate requires a resonance that also approximately lies in
the same window.
Using the value and its errors for the resonance energy published in Ref.~\cite{Ayyad:2022zqw},  $E_R=(171\pm20)$~keV, we find
\begin{align*}
	b_p &= \left(5.7_{-2.9}^{+5.0}\text{(exp.)}_{-1.1}^{+4.1}\text{(theo.)}\right)
  \times 10^{-6}\,, & 
  \log(ft)&=3.37\,,\\
	\Gamma_R &= \left(6.2^{+3.8}_{-2.6}\text{(exp.)}^{+3.9}_{-1.4}\text{(theo.)}\right)\text{~keV}\,, &
	B_{\text{GT}}&=1.63\,. 
\end{align*}
\begin{figure}[t]
  \centering
  \includegraphics[width=0.7\textwidth]{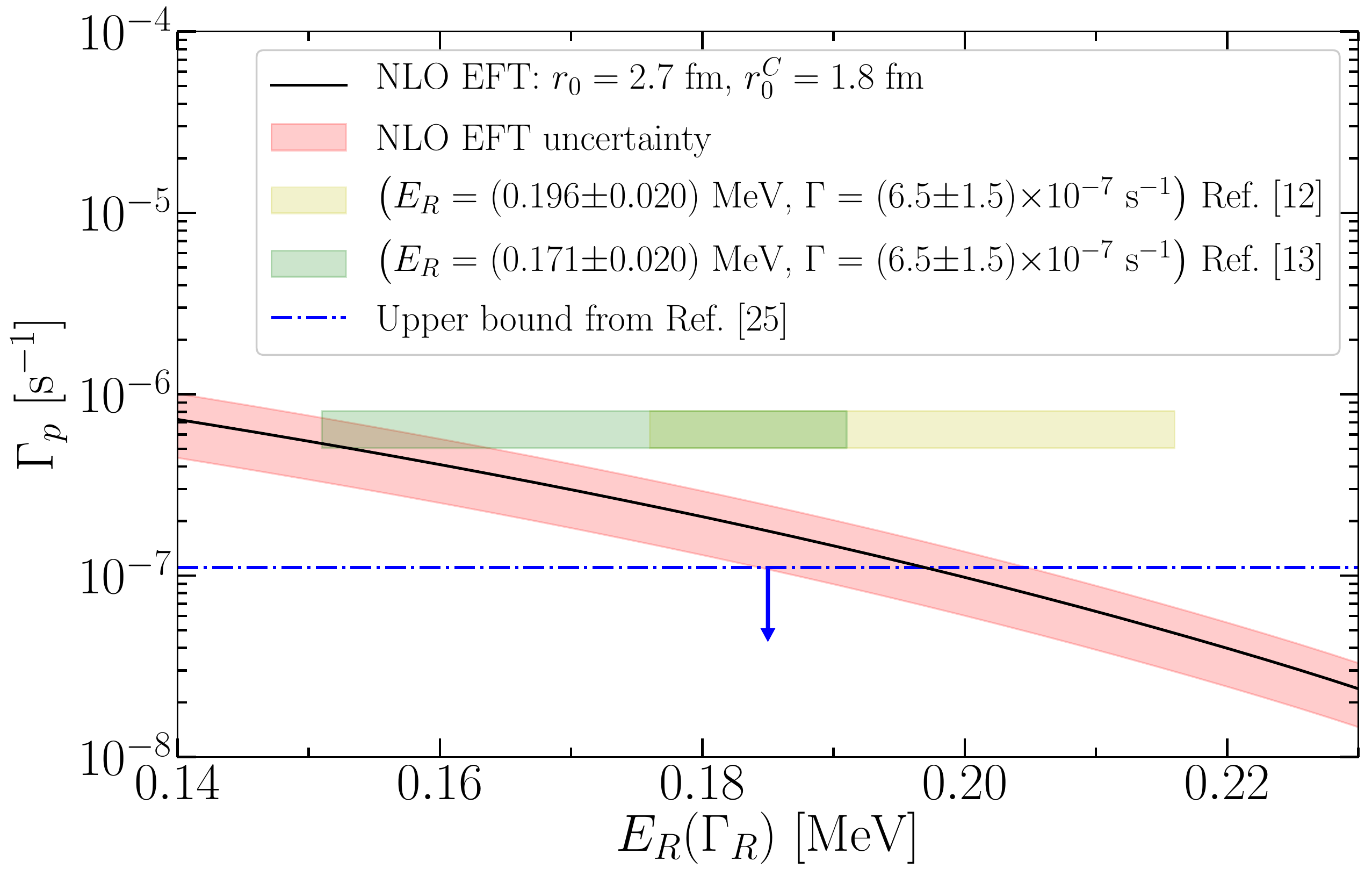}
    \caption{Partial decay rate as a function of the resonance energy
			at NLO. Explanation of 
			curves and bands is given in inset.
			}
    \label{fig:decay_rate_definite_isospin}
\end{figure}

We have also analyzed the impact of a resonance using $E_R=(196\pm 20)$~keV \cite{Ayyad:2019kna} leading to a similar plot as in Fig.~\ref{fig:diffGamma_11Be_2} with a peak position at the corresponding resonance energy.
Using the value and its errors for the resonance energy published in Ref.~\cite{Ayyad:2019kna}, we find
\begin{align*}
	b_p &= \left(2.3_{-1.3}^{+2.5}\text{(exp.)}_{-0.4}^{+1.8}\text{(theo.)}\right)
  \times 10^{-6}\,, &
  \log(ft)&=3.38\,,\\
	\Gamma_R &= \left(11.3^{+6.9}_{-4.2}\text{(exp.)}^{+7.0}_{-2.7}\text{(theo.)}\right)\text{~keV}\,, &
	B_{\text{GT}}&=1.59\,.
\end{align*}
Again both sets of values are consistent within their uncertainties.

In Fig.~\ref{fig:decay_rate_definite_isospin}, we show the partial decay rate
for beta-delayed proton emission from $^{11}$Be as a
function of the resonance energy. The solid line gives our NLO result
with the effective range parameters set as described above. The yellow and green
squares give the experimental result given in Refs.\cite{Ayyad:2019kna,Ayyad:2022zqw}, respectively. The partial decay rate decreases as the resonance
energy increases and moves out of the energy range described
above. The overlap of the green square and our EFT predictions shows that our results are consistent with the results for the decay rate published by Ayyad {\it et al.}~\cite{Ayyad:2019kna} when using the new resonance parameters determined in~\cite{Ayyad:2022zqw}. However, there is some tension when the resonance
parameters from~\cite{Ayyad:2019kna} are used. For resonance energies larger
than $\sim$~200~keV the upper bound on the partial width from
Riisager {\it et al.}~\cite{Riisager:2020glj} is satisfied. When using the resonance energies determined in Ref.~\cite{PhysRevLett.129.012502} given by $E_R=(211\pm40)$~keV leads to decay rates that are mostly in agreement with that upper bound.

\section{Results with final state interactions for Carbon-19}
\begin{figure}[t]
  \centering
  \includegraphics[width=0.7\textwidth]{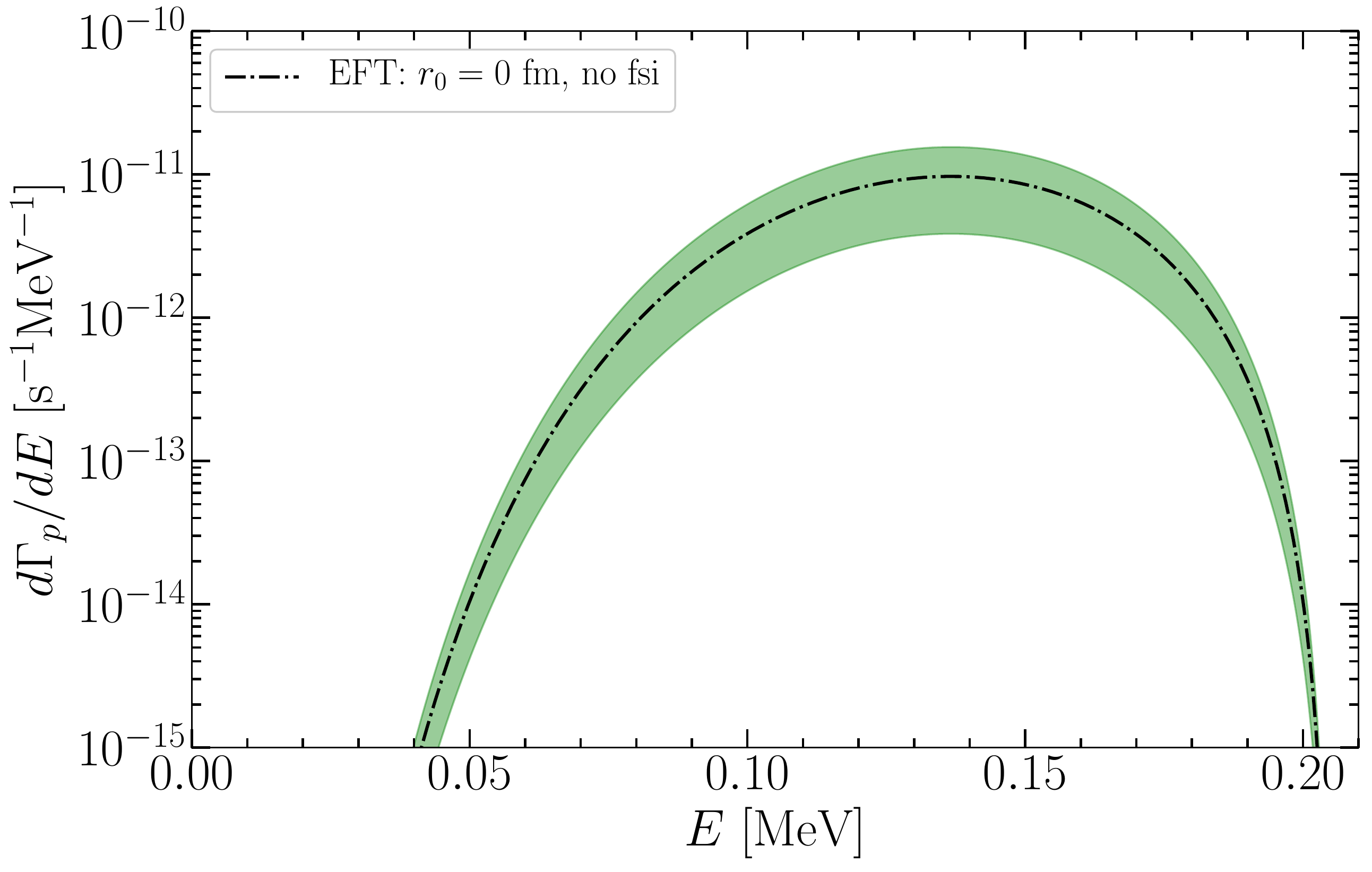}
  \caption{Differential decay rate $d\Gamma_p/dE$ for $\beta$-delayed
    proton emission from $^{19}$C as a function of the final-state
    particle energy $E$.  The dash-dotted line shows our EFT result
    without resonant final state interactions. The green band gives the EFT
    uncertainty.}
  \label{fig:diffGamma_19C}
\end{figure}
As another example, we consider beta-delayed proton emission from
the $1n$-halo nucleus $^{19}$C. In contrast to the case of $^{11}$Be,
there is no clear separation of scales between the half-lifes of
$^{19}$C which has $T_{1/2} = 46.3$~ms and of the $^{18}$C core which
has $T_{1/2} = 92\pm 2$~ms.
However, the process of beta-delayed proton emission can still be observed
by measuring the $^{18}$C core and the proton in the final state. Even if the $^{18}$C core of the $1n$-halo nucleus $^{19}$C beta decays into the ground state of $^{18}$N, it can not feed into the observed decay channel due to energy conservation since $m_{^{18}C}+m_p-(m_{^{18}N}+m_n)\approx 10.5$~MeV. We assume the contributions from decay channels via excited states of $^{18}$N$^*$ to be insignificant. 

In Fig.~\ref{fig:diffGamma_19C}, we show the differential decay rate
as a function of the energy $E$ without resonant final state interactions. 
The corresponding branching ratio is given by
\begin{align*}
	b_p &= \left(4.1\pm2.5\right)
  \times 10^{-14}\,.
\end{align*}
For comparison, the branching ratio for beta-delayed proton emission
from $^{11}$Be without resonant final state interactions is given by
$1.3 \times 10^{-8}$. Therefore, the branching ratio for $^{19}$C is
almost six orders smaller than for $^{11}$Be without final state
interactions. The reason for this is on the one hand the smaller
partial decay width due to a larger separation energy of $^{19}$C
given by $S_n=0.577$~MeV and due to the larger charge number (see
discussion in \cref{sec:results-without-final}). On the other hand,
the total decay rate of $^{19}$C is much larger leading to a further
suppression of the branching ratio according to \cref{eq:bp}. The
branching ratio could be in principle enlarged due to a $^{19}$N$^{*}$
resonance but we do not consider this effect since no such resonances
are known in the decay window.

\section{Summary}
\label{sec:summary}
In this paper, we considered $\beta$-delayed proton emission in
one-neutron halo nuclei in the framework of halo effective field
theory. We discussed the general features of the relevant matrix
elements in systems that do not have strong final state interactions
and in systems that display strong final state interactions due to a
low-lying resonance in the proton-core channel. In the scenario of no
strong final state interactions, we found that the decay rate 
predominantly depends on the one-neutron separation energy and the 
electric charge of the core. Including strong final state interactions
via a resonance in the proton-core channel can have a significant impact 
on the decay rate depending on the resonance position. If it lies within
the {\it energy window} defined by the plateau of the differential decay 
rate without strong final state interactions, it leads to a significant
enhancement of the decay rate and hence the branching ratio. We found 
that for heavier halo nuclei, a significantly enhanced branching ratio 
for a given one-neutron separation energy becomes more unlikely. The 
bigger charge of the core nucleus leads to an increased Coulomb repulsion 
between core and proton at low relative energies $E$. Thus, the 
differential decay rate for low $E$ is reduced and therefore the 
{\it energy window} for the required resonance energies in order to 
significantly enhance the branching ratio becomes smaller.
The calculations with strong final state interactions were done with and 
without isospin projection on states of definite isospin.

We have also presented additional details of the calculation for the
decay of $^{11}$Be previously published in
Ref.~\cite{Elkamhawy:2019nxq}. Furthermore, we reanalyzed our results
in the context of recently published data \cite{Ayyad:2022zqw}. We
found that the measured resonance parameters of the low-lying
resonance in the $^{11}$B system is consistent with the measured
branching ratio for the $\beta$-decay of $^{11}$Be into the continuum
but that theory and experiment do not overlap perfectly. We also
stress again that our results do not imply the existence of the
low-lying resonance but that a resonance is required for an enhanced
branching ratio of $\beta$-delayed proton emission.

Halo EFT seems generally well suited for the analysis of this
framework: Observables can be calculated in terms of a small number of
parameters that are determined from experiment. The number of
parameters used in such a calculation is also inherently tied to the
uncertainty for the observable of interest. In contradistinction to
cluster models, two-body currents appear naturally in this
framework. For example, a two-body axial current will appear one order
higher than what was considered in this work \cite{Kong:2000px}. A
higher order calculation in Halo EFT is therefore unpractical since
the unknown parameter would have to be determined from the
$\beta$-decay itself. However, this limitation is shared with other
approaches such as calculations of this process in chiral EFT with
electroweak currents. In this framework such a two-body current also
appears at next-to-next-to-leading order in the chiral
expansion~\cite{Park:1997vv}, the same order that the chiral
three-nucleon force enters which is required for the accurate
description of nuclear structure observables
\cite{Hammer:2012id}. Furthermore, we note that our approach implies
that any cluster model with the same degrees of freedom but without
additional microscopic physics will have at best the uncertainty as
our results.

There are several important questions that need to be addressed in the
future. For example, it needs to be analyzed whether a meaningful
calculation can be carried out for $P$-wave halo nuclei. In such
systems, the counting of interaction operators and current operators
changes non-trivially. It is therefore not clear whether Halo EFT or
any cluster approach can predict the lifetime accurately, however, it
provides nonetheless a complementary approach to study the decay itself. 
It would also be interesting to apply Halo EFT to
weak decays of two-neutron halo systems. Such a calculation has been
done using a cluster model~\cite{Tursunov:2017rfj}.  A halo EFT
calculation could shine light on the uncertainties of this
calculation.

\begin{acknowledgments}
  This work has been supported by the National Science Foundation
  under Grant Nos. PHY-1555030 and PHY-2111426, by the Office of
  Nuclear Physics, U.S.~Department of Energy under Contract
  No. DE-AC05-00OR22725, by the Deutsche Forschungsgemeinschaft (DFG, German 
  Research Foundation) – Projektnummer 279384907 – CRC 1245 and by the German Federal
Ministry of Education and Research (BMBF) (Grant no. 05P21RDFNB).
\end{acknowledgments}

\begin{appendix}
\section{Derivation of the decay rate}
\label{app:rate}
We choose the $z$-axis of the coordinate system in which we perform
the $\mathbf{p}_{\bar{\nu}}$-integration to be parallel to the
momentum $\mathbf{p}_e$. Therefore, the scalar product
$\pmb{\beta}_e\cdot \pmb{\beta}_{\bar{\nu}}=
|\pmb{\beta}_e|~|\pmb{\beta}_{\bar{\nu}}|
\cos\theta_{\bar{\nu}}\equiv\beta_e~\beta_{\bar{\nu}}\cos\theta_{\bar{\nu}}=\beta_e\cos\theta_{\bar{\nu}}$. Moreover,
we choose the $z$-axis of the coordinate system in which we perform
the $\mathbf{p}$-integration to be parallel to the momentum
$(\mathbf{p}_e+\mathbf{p}_{\bar{\nu}})$. We denote the angle between
$\mathbf{p}$ and $(\mathbf{p}_e+\mathbf{p}_{\bar{\nu}})$ as
$\theta$. As a consequence, the hadronic amplitude depends on
$|\mathbf{p}|$ and $|\mathbf{p}_e+\mathbf{p}_{\bar{\nu}}|$ and
$\cos\theta$. Since
\begin{align}
  |\mathbf{p}_e+\mathbf{p}_{\bar{\nu}}|=\sqrt{\mathbf{p}_e^2+\mathbf{p}_{\bar{\nu}}^2+2|\mathbf{p}_e| |\mathbf{p}_{\bar{\nu}}|\cos\theta_{\bar{\nu}}}\,,
\end{align}
the hadronic amplitude will ultimately depend on $|\mathbf{p}|$, $|\mathbf{p}_e|$, $|\mathbf{p}_{\bar{\nu}}|$, $\cos\theta$ and $\cos\theta_{\bar{\nu}}$ and therefore we replace
\begin{align}
  |\mathcal{A}\left(\mathbf{p},-\left(\mathbf{p}_e+\mathbf{p}_{\bar{\nu}}\right)\right)|^2 ~~ \rightarrow ~~ |\mathcal{A}\left(|\mathbf{p}|,|\mathbf{p}_e|,|\mathbf{p}_{\bar{\nu}}|,\cos\theta,\cos\theta_{\bar{\nu}}\right)|^2
\end{align}
and find
\begin{multline}
  \Gamma_p=\frac{G_F^2}{(2\pi)^8} \int d^3p \int d^3p_e \int d^3p_{\bar{\nu}}
  ~ \delta\left(S_n- \Delta m + \frac{\mathbf{p}^2}{2m_R} + \sqrt{\mathbf{p}_e^2+m_e^2} + |\mathbf{p}_{\bar{\nu}}|\right)\\
  \times \Big((1+3g_A^2)+\beta_e \cos\theta_{\bar{\nu}} (1-g_A^2)\Big) ~ |\mathcal{A}\left(|\mathbf{p}|,|\mathbf{p}_e|,|\mathbf{p}_{\bar{\nu}}|,\cos\theta,\cos\theta_{\bar{\nu}}\right)|^2 ~  C^2(\eta_e)\,.
\end{multline}
Now we perform the trivial angle integrations, meaning
\begin{align}
  \int d^3p ~~ &\rightarrow ~~ 2\pi \int_0^{\infty} dp ~ \mathbf{p}^2 \int_{-1}^1 d\cos\theta\,,\\
  \int d^3p_e ~~ &\rightarrow ~~ 4\pi \int_0^{\infty} dp_e ~ \mathbf{p}_e^2\,,\\
  \int d^3p_{\bar{\nu}} ~~ &\rightarrow ~~ 2\pi \int_0^{\infty} dp_{\bar{\nu}} ~ \mathbf{p}_{\bar{\nu}}^2 \int_{-1}^1 d\cos\theta_{\bar{\nu}}
\end{align}
and get 
\begin{multline}
  \Gamma_p=\frac{G_F^2}{16\pi^5} \int_0^{\infty} dp ~ \mathbf{p}^2 \int_{-1}^1 d\cos\theta \int_0^{\infty} dp_e ~ \mathbf{p}_e^2 \int_0^{\infty} dp_{\bar{\nu}} ~ \mathbf{p}_{\bar{\nu}}^2 \int_{-1}^1 d\cos\theta_{\bar{\nu}}\\
  \times \delta\left(S_n- \Delta m + \frac{\mathbf{p}^2}{2m_R} + \sqrt{\mathbf{p}_e^2+m_e^2} + |\mathbf{p}_{\bar{\nu}}|\right) \Big((1+3g_A^2)+\beta_e \cos\theta_{\bar{\nu}} (1-g_A^2)\Big)\\
  \times |\mathcal{A}\left(|\mathbf{p}|,|\mathbf{p}_e|,|\mathbf{p}_{\bar{\nu}}|,\cos\theta,\cos\theta_{\bar{\nu}}\right)|^2 ~  C^2(\eta_e)\,.
  \label{eq:B7}
\end{multline}
Using the energy-conserving delta-distribution
\begin{align}
  |\mathbf{p}_{\bar{\nu}}| ~~ \rightarrow ~~ \left(\Delta m - S_n - \frac{\mathbf{p}^2}{2m_R} - \sqrt{\mathbf{p}_e^2+m_e^2}\right)\equiv |\overline{\mathbf{p}}_{\bar{\nu}}|\,,
\end{align}
we perform the $p_{\bar{\nu}}$-integration to find
\begin{multline}
  \Gamma_p=\frac{G_F^2}{16\pi^5} \int_0^{\infty} dp \int_0^{\infty} dp_e \int_{-1}^1 d\cos\theta \int_{-1}^1 d\cos\theta_{\bar{\nu}} ~ \mathbf{p}^2 ~ \mathbf{p}_e^2 ~ |\overline{\mathbf{p}}_{\bar{\nu}}|^2 ~ \Theta\left(|\overline{\mathbf{p}}_{\bar{\nu}}|\right)\\
  \times \Big((1+3g_A^2)+\beta_e \cos\theta_{\bar{\nu}} (1-g_A^2)\Big)
  |\mathcal{A}\left(|\mathbf{p}|,|\mathbf{p}_e|,|\overline{\mathbf{p}}_{\bar{\nu}}|,\cos\theta,\cos\theta_{\bar{\nu}}\right)|^2 ~  C^2(\eta_e)\,,
\end{multline}
where $\Theta$ is the Heaviside step function.
Since $|\overline{\mathbf{p}}_{\bar{\nu}}|$ depends on $\mathbf{p}$ and $\mathbf{p}_e$, we replace
\begin{align}
  |\mathcal{A}\left(|\mathbf{p}|,|\mathbf{p}_e|,|\overline{\mathbf{p}}_{\bar{\nu}}|,\cos\theta,\cos\theta_{\bar{\nu}}\right)|^2 ~~ \rightarrow ~~ |\mathcal{A}\left(|\mathbf{p}|,|\mathbf{p}_e|,\cos\theta,\cos\theta_{\bar{\nu}}\right)|^2
\end{align}
and get
\begin{multline}
  \Gamma_p=\frac{G_F^2}{16\pi^5} \int_0^{\infty} dp \int_0^{\infty} dp_e \int_{-1}^1 d\cos\theta \int_{-1}^1 d\cos\theta_{\bar{\nu}} ~ \mathbf{p}^2 ~ \mathbf{p}_e^2 ~ |\overline{\mathbf{p}}_{\bar{\nu}}|^2 ~ \Theta\left(|\overline{\mathbf{p}}_{\bar{\nu}}|\right)\\
  \times \Big((1+3g_A^2)+\beta_e \cos\theta_{\bar{\nu}} (1-g_A^2)\Big)
  |\mathcal{A}\left(|\mathbf{p}|,|\mathbf{p}_e|,\cos\theta,\cos\theta_{\bar{\nu}}\right)|^2 ~  C^2(\eta_e)\,.
\end{multline}
Substituting 
\begin{align}
  \int_0^{\infty}dp ~ \mathbf{p}^2 &= \int_0^{\infty} dE ~ m_R \sqrt{2m_R E}\,,\\
  \int_0^{\infty}dp_e ~ \mathbf{p}_e^2 &= \int_{m_e}^{\infty} dE_e ~ E_e \sqrt{E_e^2-m_e^2}\,,
\end{align}
where $E=\mathbf{p}^2/(2m_R)$ is the relative energy of the core-proton system and replacing 
\begin{align}
  |\mathcal{A}\left(|\mathbf{p}|,|\mathbf{p}_e|,\cos\theta,\cos\theta_{\bar{\nu}}\right)|^2 ~~ \rightarrow ~~ |\mathcal{A}\left(E,E_e,\cos\theta,\cos\theta_{\bar{\nu}}\right)|^2
\end{align}
we find
\begin{multline}
  \Gamma_p=\frac{G_F^2}{16\pi^5} \int_0^{\infty} dE \int_{m_e}^{\infty} dE_e \int_{-1}^1 d\cos\theta \int_{-1}^1 d\cos\theta_{\bar{\nu}} ~ m_R \sqrt{2m_R E} ~ E_e \sqrt{E_e^2-m_e^2}\\
  \times \left(\Delta m - S_n -E -E_e\right)^2 \Theta\left(\Delta m - S_n -E -E_e\right) \Big((1+3g_A^2)+\beta_e \cos\theta_{\bar{\nu}} (1-g_A^2)\Big)\\
  \times
  |\mathcal{A}\left(E,E_e,\cos\theta,\cos\theta_{\bar{\nu}}\right)|^2 ~  C^2(\eta_e)\,.
\end{multline}

Finally, we apply the $\Theta$-function by modifying the integration areas
\begin{multline}
  \Gamma_p=\frac{G_F^2}{16\pi^5} \int_0^{\Delta m -S_n-m_e} dE \int_{m_e}^{\Delta m -S_n -E} dE_e \int_{-1}^1 d\cos\theta \int_{-1}^1 d\cos\theta_{\bar{\nu}} ~ m_R \sqrt{2m_R E}\\
  \times E_e ~ \sqrt{E_e^2-m_e^2} ~ \left(\Delta m - S_n -E -E_e\right)^2 \Big((1+3g_A^2)+\beta_e \cos\theta_{\bar{\nu}} (1-g_A^2)\Big)\\
  \times |\mathcal{A}\left(E,E_e,\cos\theta,\cos\theta_{\bar{\nu}}\right)|^2 ~  C^2(\eta_e)\,.
\end{multline}
If we do not neglect recoil effects, then there is an additional term $(\mathbf{p}_e+\mathbf{p}_{\bar{\nu}}^2)/(2M_{pc})$ in the energy-conserving delta-distribution of \cref{eq:B7}. Performing a similar calculation as before, we find for the decay rate
\begin{equation}
  \begin{aligned}
    \Gamma_p=&\frac{G_F^2 M_{pc}}{16\pi^5} \int_0^{\Delta m - S_n -m_e} dE ~ m_R \sqrt{2m_R E} \int_{-1}^1 d\cos\theta \int_{m_e}^{E_e^{\text{upper}}} dE_e ~ E_e \int_{E_{\bar{\nu}}^{\text{lower}}}^{E_{\bar{\nu}}^{\text{upper}}} dE_{\bar{\nu}} ~ E_{\bar{\nu}}  \\
    &\times \Big((1+3g_A^2)+\beta_e \overline{\cos\theta_{\bar{\nu}}} (1-g_A^2)\Big) ~ |\mathcal{A}\left(E,\cos\theta,E_e,E_{\bar{\nu}},\overline{\cos\theta_{\bar{\nu}}}\right)|^2 ~  C^2(\eta_e)\,,
  \end{aligned}
\end{equation}
with
\begin{multline}
  E_{\bar{\nu}}^{\text{lower}}=\sqrt{\left(\Delta m-S_n-E-E_e\right)2M_{pc}-(E_e^2-m_e^2)+(M_{pc}+\sqrt{E_e^2-m_e^2})^2}\\-(M_{pc}+\sqrt{E_e^2-m_e^2})\,,
\end{multline}
\begin{multline}
  E_{\bar{\nu}}^{\text{upper}}=\sqrt{\left(\Delta m-S_n-E-E_e\right)2M_{pc}-(E_e^2-m_e^2)+(M_{pc}-\sqrt{E_e^2-m_e^2})^2}\\-(M_{pc}-\sqrt{E_e^2-m_e^2})\,.
\end{multline}
and
\begin{align}
  E_e^{\text{lower}}&=m_e\,,\\
  E_e^{\text{upper}}&=\sqrt{\left(\Delta m -S_n -E\right)2M_{pc}+m_e^2+M_{pc}^2}-M_{pc}\,.
\end{align}

\end{appendix}

\end{document}